\documentclass[aps,prl,preprint,groupedaddress]{revtex4-1}

\usepackage{theorem,amsmath,amssymb}

\newcommand{\R}{{\mathbb R}}

\newtheorem{theorem}{Theorem}
\theoremstyle{theorem}
\newtheorem{proposition}[theorem]{Proposition}

\newtheorem{corollary}[theorem]{Corollary}
\theoremstyle{remark}

\begin{document}


\title{Density Functionals in the Presence of Magnetic Field}


\author{Andre Laestadius}
\email{andrela@math.kth.se}
\thanks{Department of Mathematics, KTH, 100 44 Stockholm, Sweden}
\affiliation{Department of Mathematics, KTH, 100 44 Stockholm, Sweden}


\date{\today}

\begin{abstract}
In this paper density functionals for Coulomb 
systems subjected to electric and magnetic fields are developed. The density functionals 
depend on the particle density, $\rho$, and paramagnetic current density, $j^p$. 
This approach is motivated by an adapted version of the Vignale and 
Rasolt formulation of Current Density Functional Theory (CDFT), which 
establishes a one-to-one correspondence between the non-degenerate 
ground-state and the particle and paramagnetic current density. 
Definition of $N$-representable density pairs $(\rho,j^p)$ is given and it is proven that the set of 
$v$-representable densities constitutes a proper subset of the set of $N$-representable densities. 
For a Levy-Lieb type functional $Q(\rho,j^p)$, it is demonstrated that (i) 
it is a proper extension of the universal Hohenberg-Kohn functional, $F_{HK}(\rho,j^p)$, to $N$-representable densities, 
(ii) there exists a wavefunction $\psi_0$ such that 
$Q(\rho,j^p)=(\psi_0,H_0\psi_0)_{L^2}$, where $H_0$ is the Hamiltonian 
without external potential terms, and (iii) it is not convex. Furthermore, a convex and universal functional 
$F(\rho,j^p)$ is studied and proven to be equal the convex envelope of $Q(\rho,j^p)$. For both $Q$ and $F$, we give upper and lower bounds. 
\end{abstract}


\maketitle

\section{I. INTRODUCTION}
The theoretical foundation of Density Functional Theory (DFT) is the Hohenberg-Kohn theorem \cite{HK64} that states that the particle density 
of a quantum mechanical system determines the scalar potential up to a constant. Arguments have been put forward that this theorem could be 
generalized to include systems with magnetic fields \cite{V87,Diener,Sahni}. These arguments rely on 
either the paramagnetic current density or the total current density being used together 
with the particle density to determine the scalar potential and vector potential of the system. Nonetheless, for the formulation with 
paramagnetic current density, counterexamples have been constructed that exclude the existence of a Hohenberg-Kohn theorem for such 
a formulation (see for instance \cite{CV2002} where this was first 
demonstrated, or \cite{AndreMichael} for more mathematical details in the one-electron case). Moreover, the existence of a 
Hohenberg-Kohn theorem for the formulation with the total current density 
is still an open question \cite{AndreMichael,Tellgren}, since the proofs of \cite{Diener} and \cite{Sahni} do not hold. 
(The error in \cite{Diener} was highlighted in \cite{AndreMichael} and the error in \cite{Sahni} was pointed out in \cite{Tellgren}.)

However, in an adapted version of the Vignale and Rasolt formulation of Current Density Functional Theory (CDFT), the particle density and 
the paramagnetic current density determine the non-degenerate ground-state \cite{V87,AndreMichael}. This allows a Hohenberg-Kohn 
functional to be defined, from which other density functionals can be developed. Following Lieb's programme for DFT \cite{Lieb83}, 
the issue of establishing a mathematically rigorous CDFT formulated with the paramagnetic current density will here be addressed.\\

The aims of this article are the following: 

(i) {\it Define the set of $N$-representable particle and paramagnetic current densities.} The 
definition is motivated by Proposition 3, and Proposition 4 
shows that this set is convex.

(ii) {\it For $N$-representable particle and paramagnetic current densities, study 
a Levy-Lieb type functional $Q(\rho,j^p)$.} In Theorem 5, 
$Q(\rho,j^p)$ is proven to be a proper extension of the universal Hohenberg-Kohn functional and, moreover, it is proven that 
there exists a minimizer such that 
$Q(\rho,j^p)=(\psi_0,H_0\psi_0)_{L^2}$. 
Proposition 8 demonstrates that $Q(\rho,j^p)$ is not a convex functional, 
which motivates 

(iii) {\it Investigate a convex and universal particle and paramagnetic current 
density functional, $F(\rho,j^p)$.} In Theorem \ref{ConvexEn}, it is proven that 
$F(\rho,j^p)$ equals the convex envelope of $Q(\rho,j^p)$. Furthermore, in Theorem \ref{thmB}, the minimization of 
$$F(\rho,j^p) + 2\int_{\R^3}j^p\cdot A + \int_{\R^3}\rho(v + |A|^2)$$ is connected with a set of Euler-Lagrange equations.

(iv) {\it Give upper and lower bounds for particle and paramagnetic current density functionals.} Bounds for both $Q(\rho,j^p)$ and $F(\rho,j^p)$ are 
found in Theorem \ref{GEN3.8}, Proposition \ref{propD} and Corollary \ref{corE}. 
Proposition \ref{propD} and Corollary \ref{corE} require that the vorticity is zero.

\section{II. PRELIMINARIES}
We will in this paper consider a system of $N$ interacting electrons. The Hamiltonian of the system is given by (in suitable units)
\begin{align}
H(v,A)=\sum_{k=1}^N \left((i\nabla_k - A(x_k))^2 + v(x_k) \right) + \sum_{1\leq k<l\leq N} |x_k - x_l|^{-1},
\label{Hamil}
\end{align}
where $v(x)$ is the scalar potential and $A(x)$ the vector potential, with components $A^k(x)$ $k=1,2,3$, such that $B(x)= \nabla \times A(x)$, 
where $B(x)$ is the magnetic field. The following will be assumed: (i) there is a lowest eigenvalue $e_0$ of $H(v,A)$ with 
$\text{dim\,ker}(e_0 -H)=1$, (ii) the solution of $H(v,A)\psi =e_0\psi$ fulfils $\psi\neq 0$ almost everywhere (a.e.), and 
(iii) the magnetic field vanishes outside some large sphere ($B$ has compact support) and we may take $A^k(x)$ to be bounded. 
See \cite{AndreMichael} for further discussion about assumptions (i) and (ii).

Some different function spaces will be used in the forthcoming discussion. A function $f$ that satisfies 
$\int_{\R^n} |f|^p<\infty$, for some $p\in [1,\infty)$, is said to belong to the normed space $L^p(\R^n)$ with 
norm $||f||_{L^p(\R^n)} = \left(\int_{\R^n} |f|^p\right)^{1/p}$. For $R>0$, let $B_R = \{x\in\R^n| \,|x|\leq R \}$. Then 
$f\in L_{\text{loc}}^p(\R^n)$ if for any $B_R$ we have that $||f||_{L^p(B_R)} =\left(\int_{B_R} |f|^p\right)^{1/p}<\infty$. The normed space 
$L^\infty(\R^n)$ consists of those functions $f$ that satisfy $||f||_{L^\infty(\R^n)} = \text{ess} \sup \{|f|\, |x\in \R^n \}<\infty$. 
A function $f\in L^2(\R^n)$ that satisfies $\int_{\R^n} |\nabla f|^2<\infty$ belongs to $H^1(\R^n)$. Furthermore, 
$f\in L^2(\R^n)$ belongs to $H_A^1(\R^n)$ if 
$\int_{\R^n} |(i\nabla - A)f|^2 <\infty$ for some non-zero $A(x)$. Both $H^1$ and $H_A^1$ are Hilbert spaces with norms 
$|| f||_{H^1}^2 = \int_{\R^n}|f|^2 + \int_{\R^n}|\nabla f|^2$ and 
$|| f||_{H_A^1}^2 = \int_{\R^n}|f|^2 + \int_{\R^n}|(i\nabla -A)f|^2$ respectively. 
For a vector $u$, if each component of $u$, $(u)_l$, $l=1,2,3$, belongs to $L^p$ for some $p\in [1,\infty]$, we write $u\in (L^p)^3$.

For the proofs set forth in this article, some different notions of convergence will be used. 
A sequence $\{\psi_k\}\subset L^p(\R^n)$ is said to converge 
(in $L^p$-norm) to $\psi \in L^p(\R^n)$ if and only if $\int_{\R^n} |\psi_k - \psi|^p \rightarrow 0$, and 
we write $\psi_k\rightarrow \psi$. Moreover, denote the inner product of a Hilbert space $H$ by $(\cdot,\cdot)_H$. 
A sequence $\{\psi_k \}\subset H$ is then said to converge weakly to $\psi\in H$ if and only if 
$(\psi_k,\phi)_H \rightarrow (\psi,\phi)_H$ for all $\phi\in H$, and we write $\psi_k\rightharpoonup \psi$. 
The inner product of $H^1(\R^n)$ is given by 
$(\psi,\phi)_{H^1(\R^n)} = \int_{\R^n }\overline{\psi} \phi +  \int_{\R^n }\overline{\nabla \psi}\cdot \nabla \phi$. In particular, 
weak convergence on $H^1(\R^n)$ implies weak convergence in the $L^2(\R^n)$ sense, i.e., 
$(\psi_k,\phi)_{L^2(\R^n)}= \int_{\R^n}\overline{\psi_k} \phi \rightarrow \int_{\R^n }\overline{\psi} \phi =(\psi,\phi)_{L^2(\R^n)}$.

Also note that a function (or functional) $f: D \rightarrow \R$ is convex on $D$ if for $x_1,x_2\in D$ and $0\leq \lambda\leq 1$, 
we have $f(\lambda x_1 + (1-\lambda)x_2)\leq \lambda f(x_1) + (1-\lambda)f(x_2)$.\\

Now, let $\psi$ denote the wavefunction describing the system. For simplicity, spin will not be treated. Henceforth, assume that 
$\psi(x_1,\dots,x_N)$ is antisymmetric in its coordinates $x_i$ and belongs to 
\begin{align}
W_{N}= \{\psi\in H^1(\R^{3N})| \,||\psi||_{L^2(\R^{3N})}=1 \}.
\label{WWW}
\end{align}
Assume $A^k\in L^\infty(\R^3)$, $k=1,2,3$, and $v\in L^{3/2}(\R^3) + L^\infty(\R^3)$, and define 
the ground-state energy
\begin{align}
e_0(v,A)= \inf\left\{E_{v,A}(\psi) | \psi \in W_{N} \right\},
\label{GSenergy}
\end{align}
where $E_{v,A}(\psi)$ is a functional on $W_{N}$ given by
\begin{align}
E_{v,A}(\psi)= \sum_k \left(\int_{\R^{3N}} |(i\nabla_k - A(x_k))\psi|^2 + \int_{\R^{3N}} |\psi|^2v(x_k) \right)
+\sum_{k<l} \int_{\R^{3N}} |\psi|^2|x_k - x_l|^{-1}.
\label{Schr}
\end{align}
We shall interpret the inner-product $(\psi,H(v,A)\psi)_{L^2}$ as the number $E_{v,A}(\psi)$, which is well-defined for $\psi\in W_{N}$.

For $\psi\in W_N$, define the particle density and 
the paramagnetic current density to be, respectively,
\begin{align}
\rho_\psi(x) &=N \int_{\R^{3(N-1)}} |\psi(x,x_2,\dots,x_N)|^2 dx_2\dots dx_N,\nonumber \\
j_\psi^p(x) &=N \,\text{Im} \int_{\R^{3(N-1)}} \overline{\psi}(x,x_2,\dots,x_N)\nabla_x \psi (x,x_2,\dots,x_N)dx_2\dots dx_N.
\label{Densities}
\end{align}
Let $H(v,A)$ for the special case $v=0$ and $A=0$ be denoted $H_0$, that is,
\begin{align*}
H_0=-\sum_{k=1}^N \Delta_k + \sum_{1\leq k<l\leq N} |x_k - x_l|^{-1}, 
\end{align*}
and set 
\begin{align}
(\psi,H_0\psi)_{L^2} = \sum_k \int_{\R^{3N}} |\nabla_k \psi|^2 dx_1\dots dx_N +\sum_{k<l} \int_{\R^{3N}} |\psi|^2|x_k - x_l|^{-1}dx_1\dots dx_N,
\label{H0}
\end{align}
for $\psi\in W_N$, even though $H_0\psi \notin L^2$. The kinetic energy of $\psi$, denoted $T(\psi)$, and 
the exchange-correlation energy, denoted $E_{xc}(\psi)$, are given by, respectively,
\begin{align*}
T(\psi) &= \sum_{k=1}^N \int_{\R^{3N}}|\nabla_k \psi|^2dx_1\dots dx_N,\\
E_{xc}(\psi) &= (\psi,\sum_{1\leq k<l\leq N}|x_k-x_l|^{-1}\psi)_{L^2}- 
\frac{1}{2} \int_{\R^3} \int_{\R^3} \frac{\rho_\psi(x)\rho_\psi(y)}{|x-y|}dx dy.
\end{align*}
Note that \eqref{H0} can be written as $(\psi,H_0\psi)_{L^2} = T(\psi) + E_{xc}(\psi) + 
\frac{1}{2} \int_{\R^3} \int_{\R^3} \frac{\rho_\psi(x)\rho_\psi(y)}{|x-y|}dxdy$.\\

To put this work into context, the case $A=0$ will first be discussed. A particle density $\rho$ is said to be $v$-representable if there exists a Hamiltonian 
$H(v)$, with ground-state $\psi_0$, such that $\rho = \rho_{\psi_0}$. The Hohenberg-Kohn theorem then states that a $v$-representable 
particle density $\rho$ determines the scalar potential $v(x)$ up to a constant \cite{HK64}. For such densities, we can define
\begin{align*}
F_{HK}(\rho) = (\psi_\rho, H(v_\rho)\psi_\rho)_{L^2} - \int_{\R^3} \rho v_\rho=(\psi_\rho, H_0\psi_\rho)_{L^2},
\end{align*}
where $\psi_\rho$ is the ground-state of $H(v_\rho)$ and where $v_\rho$ is determined by $\rho$ (according to the Hohenberg-Kohn theorem). 
This scheme, however, suffers from the fact that the functional $F_{HK}(\rho)$ is not explicitly computable, and that the set of 
$v$-representable particle densities is unknown. To remedy this situation, Lieb \cite{Lieb83} extended the Hohenberg-Kohn functional $F_{HK}(\rho)$ 
to $F_{LL}(\rho)$ for $\rho\in I_N$, where
\begin{align}
F_{LL}(\rho) = \inf\{ (\psi,H_0\psi)_{L^2}|\psi\in W_N, \rho_\psi=\rho \},
\label{FLL}
\end{align}
and
\begin{align*}
I_N = \Big\{\rho\, \Big| \rho \geq 0, \rho^{1/2}\in H^1(\R^3), \int_{\R^3} \rho=N\Big\}.
\end{align*}
The ground-state energy, $e_0(v)$, can then be 
obtained from 
\begin{align*}
e_0(v) = \inf\Big\{F_{LL}(\rho) + \int_{\R^3}\rho v\,\Big|\, \rho\in I_N   \Big\},
\end{align*}
which is the so-called Levy-Lieb constrained search formalism \cite{Lieb83,Levy}. 
Moreover, Lieb \cite{Lieb83} has proved that the functional $F_{LL}(\rho)$ is not convex and that the functional
\begin{align*}
F_c(\rho) = \sup\Big\{e_0(v) - \int_{\R^3}\rho v \Big| v\in L^{3/2}(\R^3) + L^\infty(\R^3)\Big\},
\end{align*}
which is convex, equals the convex envelope of $F_{LL}(\rho)$ on $L^1(\R^3)\cap L^3(\R^3)$. Furthermore, 
\begin{align*}
e_0(v) = \inf\Big\{F_{c}(\rho) + \int_{\R^3}\rho v\,\Big|\, \rho\in   L^1(\R^3)\cap L^3(\R^3) \Big\}.
\end{align*}
This is the programme we now wish to undertake for paramagnetic current density functionals.

In the remainder of this paper the system Hamiltonian, $H(v,A)$, will also account for a magnetic field $B(x)\neq 0$. 
For such a system, both the particle density and the paramagnetic current density are needed to describe the system.

\section{III. PARAMAGNETIC CURRENT DENSITY FUNCTIONALS}
We will here begin the pursuit of describing a system of $N$ interacting electrons in terms of density functionals with both $\rho$ and 
$j^p$ as variables. 
The Vignale and Rasolt formulation of CDFT uses the paramagnetic current density $j^p$ together with the particle density $\rho$. The statement 
in \cite{V87} that the potentials $v$ and $A$ are determined by the density pair $(\rho,j^p)$ can be reformulated to correctly state that 
$(\rho,j^p)$ determines the non-degenerate ground-state wavefunction $\psi$. This ground-state $\psi$ may be the solution 
to many different Schr\"odinger equations of the 
form $H(v,A)\psi =e_0\psi$, where $e_0$ is the lowest eigenvalue and $\psi$ assumed to be non-degenerate. Hence, the density pair $(\rho,j^p)$ does not necessarily determine the potentials 
$v$ and $A$.

With this correspondence between a density pair $(\rho,j^p)$ and a non-degenerate ground-state $\psi$, the aim 
is now to generalize some previous results for particle density functionals, i.e., functionals that only depend on $\rho$. This generalization will follow Lieb's programme for DFT 
\cite{Lieb83}, and will constitute of the following: 
(i) give mathematical criteria for $N$-representable density pairs $(\rho,j^p)$, (ii) extend a universal Hohenberg-Kohn functional, 
denoted $F_{HK}(\rho,j^p)$, to a Levy-Lieb-type functional, denoted $Q(\rho,j^p)$, which has the $N$-representable densities as domain, 
(iii) study the convex envelope of $Q(\rho,j^p)$, and (iv) give upper and lower bounds for both $Q(\rho,j^p)$ and $F(\rho,j^p)$.

\subsection{A. The Hohenberg-Kohn functional and $N$-representable densities}
The starting point is the following theorem:
\begin{theorem}
Assume that $H(v_1,A_1)$ and $H(v_2,A_2)$ have non-degenerate ground-states $\psi$ and $\phi$ respectively. 
Then $\rho_\psi=\rho_\phi$ and $j_\psi^p=j_\phi^p$ imply $\psi = \text{const.}\,\phi$.
\end{theorem}
\textbf{Remarks.} (i) For a proof we refer either to \cite{V87} or Theorem 9 in \cite{AndreMichael}.

(ii) Note that Theorem 1 differs from the Hohenberg-Kohn theorem \cite{HK64} since no claim is made that the densities determine the potentials.\\

Theorem 1 will now be applied. The first issue to address is a Hohenberg-Kohn functional that depends on both the density $\rho$ and 
the paramagnetic current density $j^p$.\\
\\
\textbf{Definition.} A density pair $(\rho,j^p)$ is said to be $v$-representable if there exists a Hamiltonian $H(v,A)$, with ground-state 
$\psi_0$, such that $\rho=\rho_{\psi_0}$ and $j^p = j_{\psi_0}^p$. This set of densities will be denoted $A_N$, that is,
$$A_N = \{(\rho,j^p)\vert \rho=\rho_\psi, j^p=j_\psi^p, \text{$\psi$ is a non-degenerate ground-state of 
some $H(v,A)$} \}.$$
\\
\indent For $(\rho,j^p)\in A_N$, let $\psi_{\rho,j^p}$ denote the non-degenerate ground-state of 
some $H(v,A)$, which is determined by $(\rho,j^p)$ according to Theorem 1. The Hohenberg-Kohn functional, given by
\begin{align*}
F_{HK}(\rho,j^p) = (\psi_{\rho,j^p},H_0 \psi_{\rho,j^p})_{L^2},
\end{align*}
is then well-defined for $(\rho,j^p)\in A_N$. Let the set of those potentials $v$ and $A$ such that $H(v,A)$ has a non-degenerate ground-state 
be denoted $V_N$, i.e.,$$V_N =\{(v,A)| \text{$H(v,A)$ has a non-degenerate ground-state}   \}.$$
For $(v,A)\in V_N$, one has
\begin{align*}
e_0(v,A) = \min\Big\{F_{HK}(\rho,j^p) + 2\int_{\R^3}j^p\cdot A + \int_{\R^3}
\rho(v+ |A|^2)  \Big|(\rho,j^p)\in A_N  \Big\}.
\end{align*}
This is the so-called variational principle of CDFT.
\begin{theorem}
For a given potential pair $(v,A)\in V_N$, the ground state energy functional assumes its minimum value for the true ground state densities if the 
admissible densities are in $A_N$, i.e., 
$$e_0(v,A)= \min \Big\{ F_{HK}(\rho,j^p) + 2\int_{\R^3}j^p\cdot A + \int_{\R^3}
\rho(v+ |A|^2)  \Big| (\rho, j^p)\in A_N  \Big\}.$$
\label{Thm1.10}
\end{theorem}
{\it Proof.} Fix $(v,A)\in V_N$. For any $(\rho,j^p) \in A_N$ there exist potentials $\tilde{v}$ and $\tilde{A}$ such that 
$H(\tilde{v},\tilde{A})$ has a non-degenerate ground-state $\psi_{\rho,j^p}$, and one can define
\begin{align*}
G_{v,A}(\rho,j^p) = ( \psi_{\rho,j^p},H(v,A)  \psi_{\rho,j^p})_{L^2}. 
\end{align*}
Since $\psi_{\rho,j^p}$ need not be the ground state of $H(v,A)$, 
by the variational principle for wavefunctions 
\begin{align*}
G_{v,A}(\rho,j^p)\geq e_0(v,A).
\end{align*}
Furthermore, by the fact that $(v,A) \in V_N$, there exists a non-degenerate ground state $\psi_0$ of $H(v,A)$. Let 
$\rho_0 = \rho_{\psi_0}$ and $j_0^p = j_{\psi_0}^p$, that is, the corresponding ground state particle and paramagnetic current density, 
which clearly belong to $A_N$. For $(\rho_0,j_0^p)\in A_N$ there exists $\psi_{\rho_0,j_0^p}$ that satisfies 
$\psi_{\rho_0,j_0^p} =\text{const.}\, \psi_0$, by Theorem 1. Hence 
\begin{align*}
G_{v,A}(\rho_0,j_0^p) =( \psi_{\rho_0,j_0^p}, H(v,A)  \psi_{\rho_0,j_0^p})_{L^2} =  e_0(v,A).
\end{align*} 
One may then conclude that
\begin{align*}
e_0(v,A) &= \min\left\{G_{v,A}(\rho,j^p)\vert (\rho,j^p)\in A_N  \right\}\\
&=\min \Big\{F_{HK}(\rho,j^p) +2 \int_{\R^3} j^p\cdot A + \int_{\R^3} \rho (v + |A|^2)  \Big| (\rho, j^p)\in A_N  \Big\}. \,\,\,\blacksquare 
\end{align*}
\\
\indent The next step is to define a Levy-Lieb-type functional. This functional will be denoted $Q(\rho,j^p)$ and will 
depend on density pairs $(\rho,j^p)$ that are said to be $N$-representable. To that end, first note
\begin{proposition}
(i) If $\psi\in W_N$, then $\rho_\psi\in I_N$, $j_\psi^p\in (L^1(\R^3))^3$ and 
$\int_{\R^3}|j_\psi^p|^2 \rho_\psi^{-1}\leq T(\psi)$, and

(ii) the functional $(\rho,j^p) \mapsto \int_{\R^3}|j^p|^2 \rho^{-1}<\infty$ is convex. 
\label{PROP1}
\end{proposition}
{\it Proof.} (i) Let $\psi\in W_N$, then by Theorem 1.1 of \cite{Lieb83}, $\rho_\psi\in I_N$. Furthermore, one has
\begin{align*}
\int_{\R^3}|j_\psi^p|^2 \rho_\psi^{-1} \leq \sum_{k=1}^3 \int_{\R^3}
\left(N^2\int_{\R^{3(N-1)}} |\psi|^2  \int_{\R^{3(N-1)}}|\partial_k\psi|^2 \right)\rho_\psi^{-1} 
= N\int_{\R^{3N}} |\nabla_1 \psi|^2 =T(\psi).
\end{align*}
To see that each component of $j_\psi^p$ is in $L^1(\R^3)$, note that 
\begin{align*}
\int_{\R^3} \left| \int_{\R^{3(N-1)}}\text{Im} (\overline{\psi} \partial_k\psi)\right|
\leq\int_{\R^{3N}}| \overline{\psi} \partial_k\psi|
\leq \left(\int_{\R^{3N}}|\psi|^2\right)^{1/2}  \left(\int_{\R^{3N}}|\partial_k\psi|^2\right)^{1/2}<\infty.
\end{align*}

To prove (ii), set $\rho = \lambda\rho_1 + (1-\lambda)\rho_2$ and $j^p=\lambda j_1^p + (1-\lambda) j_2^p$, where $0<\lambda <1$. Since 
$$\frac{\rho_2}{\rho_1}|j_1^p|^2 +  \frac{\rho_1}{\rho_2} |j_2^p|^2 \geq 2j_1^p\cdot j_2^p,$$
it follows that
\begin{align*}
\rho\left(\lambda  \frac{|j_1^p|^2}{\rho_1}  + (1-\lambda)\frac{|j_2^p|^2}{\rho_2}   \right)
&= \lambda^2 |j_1^p|^2 + \lambda(1-\lambda) \left( \frac{\rho_2}{\rho_1}|j_1^p|^2 +  \frac{\rho_1}{\rho_2}|j_2^p|^2 \right)
+ (1-\lambda)^2 |j_2^p|^2\\
&\geq \lambda^2 |j_1^p|^2 + 2\lambda(1-\lambda) j_1^p\cdot j_2^p + (1-\lambda)^2 |j_2^p|^2
=\left\vert  j^p\right\vert^2.
\end{align*}
One may then conclude 
\begin{align*}
\int_{\R^3} |j^p|^2 \rho^{-1} 
\leq \lambda\int_{\R^3} |j_1^p|^2\rho_1^{-1} +(1-\lambda)\int_{\R^3} |j_2^p|^2\rho_2^{-1},
\end{align*}
which shows the convexity. $\,\,\,\blacksquare$\\
\\
\indent Motivated by Proposition \ref{PROP1}, the set of $N$-representable density pairs $(\rho,j^p)$ is now defined as follows.\\ 
\\
\textbf{Definition.} A density pair $(\rho,j^p)$ is said to be $N$-representable if $(\rho,j^p)\in Y_N$, 
where $Y_N = \Big\{(\rho,j^p) \Big|\rho\in I_N, j^p \in (L^1(\R^3))^3, \int_{\R^3}|j^p|^2 \rho^{-1}<\infty   \Big\}$.\\

A convex combination of $N$-representable densities is also $N$-representable, but a $v$-representable density need not be 
$N$-representable. To summarize

\begin{proposition}
(i) The set $Y_N$ is convex, and \\
(ii) $A_N \subsetneq Y_N$.
\end{proposition}
\textbf{Remark.} The proof of part (ii) will be given after Proposition 8.\\

\noindent {\it Proof of (i).} Recall that the set $I_N$ consists of those non-negative densities $\rho$ that satisfy $\rho^{1/2}\in H^1(\R^3)$ and 
$\int_{\R^3}\rho = N$. Note that the functional $\rho\mapsto \int_{\R^3} (\nabla\rho^{1/2})^2$ is convex and that $I_N$ is a convex 
set \cite{Lieb83}. Since by Proposition \ref{PROP1} (ii), 
$(\rho,j^p) \mapsto \int_{\R^3}|j^p|^2 \rho^{-1}<\infty$ is a convex functional, it follows that $Y_N$ is a convex set. $\,\,\,\blacksquare$\\

Note that for $(\rho,j^p)\in Y_N$, $\int_{\R^3}\rho v$ and $\int_{\R^3}j^p\cdot A$ are finite since $v\in L^{3/2}(\R^3) + L^\infty(\R^3)$ 
and $A^k \in L^\infty(\R^3)$. However, if a given $A$ has $A^k\notin L^\infty(\R^3)$ for some $k$, $\int_{\R^3}j^p\cdot A$ 
is still finite if $(\rho,j^p)\in Y_A$, where $Y_A = \{ (\rho,j^p)\in Y_N | \rho\in L^1(\R^3,|A|^2)   \}$. Note that 
if $$\psi \in \tilde{W}_{N,A}=\{\psi\in \otimes_{k=1}^N H_A^1(\R^{3})| \psi\in W_N,\int_{\R^3}\rho_\psi |A|^2<\infty\},$$ then $(\rho,j^p)\in  Y_A$, 
and 
\begin{align}
\int_{\R^3} |(j^p)_k \, A^k| \leq \left(\int_{\R^3}|j^p|^2\rho^{-1}\right)^{1/2}
\left(\int_{\R^3}\rho |A|^2 \right)^{1/2}<\infty.
\label{Mosk1}
\end{align}
The proof of \eqref{Mosk1} follows directly from $\int_{\R^3} |(j^p)_kA^k|= \int_{\R^3} |(j^p)_k\rho^{-1/2}| |\rho^{1/2}A^k|$ and 
using Schwarz's inequality.

\subsection{B. The Levy-Lieb-type functional $Q(\rho,j^p)$}
We now turn to finding an extension of the functional $F_{HK}(\rho,j^p)$. A Levy-Lieb-type functional, denoted $Q(\rho,j^p)$, 
will be introduced (cf. \cite{Lieb83} and \cite{Levy}) and proven to satisfy $Q (\rho,j^p)=F_{HK}(\rho,j^p)$ for $(\rho,j^p)\in A_N$. 
The domain of $Q(\rho,j^p)$ will consist of those $\rho$ and $j^p$ that are elements of $Y_N$.\\
\\
\textbf{Definition.} For $(\rho,j^p)\in  Y_N$, we define a Levy-Lieb-type functional
\begin{align*}
Q(\rho,j^p) = \inf \{ (\psi,H_0 \psi)_{L^2}| \psi \in W_N,\psi\mapsto(\rho,j^p) \},
\end{align*}
where $\psi\mapsto(\rho,j^p)$ means that $\rho_\psi = \rho,j_\psi^p =j^p$.\\
\\
\indent Note that $Q(\rho,j^p)$ is the generalization of $F_{LL}(\rho)$, see \eqref{FLL}, when also describing the system with the paramagnetic 
current density. Theorem 3.3 of \cite{Lieb83} states that there exists a $\psi_0\in H^1(\R^{3N})$ such that 
$F_{LL}(\rho) = (\psi_0,H_0\psi_0)_{L^2}$ and $\psi_0\mapsto \rho$ for $\rho\in I_N$. 
A similar result is also true for the functional $Q(\rho,j^p)$. Furthermore, on $Y_N$, $Q(\rho,j^p)\geq\int_{\R^3}|j^p|^2\rho^{-1}$. 
The next theorem summarizes the claims made so far about $Q(\rho,j^p)$. 
\begin{theorem}
(i) There exists a $\psi_0$ such that $Q(\rho,j^p) = (\psi_0,H_0 \psi_0)_{L^2}$ and $\psi_0\mapsto (\rho,j^p)$,\\
(ii) $Q(\rho,j^p)$ is the proper extension of $F_{HK}(\rho,j^p)$ from $A_N$ to $Y_N$ in the sense 
that for $(\rho,j^p)\in A_N$, $Q(\rho,j^p)=F_{HK}(\rho,j^p)$, and\\
(iii) $\int_{\R^3}|j^p|^2\rho^{-1}\leq Q(\rho,j^p)$ on $Y_N$.
\label{thm3.4}
\end{theorem}

\noindent {\it Proof.}
(i) Let $\{\psi^k\}_{k=1}^\infty$ be a minimizing sequence, that is 
$\lim_k (\psi^k,H_0 \psi^k)_{L^2} = Q(\rho,j^p)$ and $\psi^k\mapsto (\rho,j^p)$ for all $k$. 
From \cite{Lieb83} (Theorem 3.3), 
$\psi^k \rightharpoonup \psi_0$ in 
$H^1(\R^{3N})$ and $\psi^k \rightarrow \psi_0$ in $L^2(\R^{3N})$ for some 
$H^1$-function $\psi_0$ (after passing to a subsequence, which we for simplicity continue to denote $\psi^k$). 
Then by Theorem 1.3 and Theorem 3.3 in \cite{Lieb83}, $\rho_{\psi_0}=\rho$. 
Since taking weak limits, one has $\lim_k (\psi^k,H_0 \psi^k)_{L^2}\geq (\psi_0,H_0 \psi_0)_{L^2}$. It remains to show 
that $\psi_0\mapsto j^p$ a.e.

Let $g(x)=\chi_M(x)$ be the characteristic function of any (measurable) set $M\subset \R^3$ and 
let $(u)_l$ denote the $l$:th component of the vector $u$. 
Now, using the weak convergence of $\{ \psi^k\}_{k=1}^\infty$ in $H^1(\R^{3N})$ and the norm-convergence in $L^2(\R^{3N})$, we have 
for $l=1,2,3$,
\begin{align*}
\lim_{k\rightarrow \infty}\int_{\R^3}(j_{\psi^k}^p)_l g &= \lim_{k\rightarrow \infty} N\,\text{Im}\,\int_{\R^3} \int_{\R^{3(N-1)}} \overline{\psi^k} (\partial_l \psi^k) g \\
&= \lim_{k\rightarrow \infty}N\,\text{Im}\,\left(\int_{\R^{3N}} (\overline{\psi^k} -\overline{\psi_0})(\partial_l \psi_k) g + 
\int_{\R^{3N}} \overline{\psi_0}(\partial_l \psi^k) g  \right)
= \int_{\R^3}(j_{\psi_0}^p)_l g.
\end{align*}
This gives $j_{\psi_0}^p(x) =j_{\psi^k}^p(x) = j^p(x)$ a.e.

(ii) Fix $(\rho,j^p)\in A_N$ and let $\psi_0$ be as in part (i). The claim in (ii) will be shown by demonstrating that $(\psi_0,H_0\psi_0)_{L^2} =(\psi_{\rho,j^p},H_0\psi_{\rho,j^p})_{L^2}$, 
where $\psi_{\rho,j^p}$ satisfies $F_{HK}(\rho,j^p) = (\psi_{\rho,j^p},H_0\psi_{\rho,j^p})_{L^2}$. Now, since $\psi_{\rho,j^p}\in W_N$ and 
$\psi_{\rho,j^p}\mapsto (\rho,j^p)$, $$(\psi_0,H_0\psi_0)_{L^2} = Q(\rho,j^p)\leq (\psi_{\rho,j^p},H_0\psi_{\rho,j^p})_{L^2}.$$ On the other hand, 
since $\psi_{\rho,j^p}$ is the ground state of some Hamiltonian $H(v,A)$, 
\begin{align*}
e_0(v,A) &= (\psi_{\rho,j^p},H(v,A)\psi_{\rho,j^p})_{L^2} = (\psi_{\rho,j^p},H_0\psi_{\rho,j^p})_{L^2}  + 2\int_{\R^3}j^p\cdot A +\int_{\R^3}\rho(v+|A|^2)\\
&\leq (\psi_0,H(v,A)\psi_0)_{L^2} =(\psi_0,H_0\psi_0)_{L^2} + 2\int_{\R^3}j^p\cdot A +\int_{\R^3}\rho(v+|A|^2),
\end{align*}
and $(\psi_0,H_0\psi_0)_{L^2} \geq (\psi_{\rho,j^p},H_0\psi_{\rho,j^p})_{L^2}$.

(iii) Fix $(\rho,j^p)\in Y_N$. Let $\psi\in W_N$ such that $\rho_\psi =\rho$ and $j_\psi^p = j^p$. By Proposition \ref{PROP1} (i), 
$\int_{\R^3}|j^p|^2\rho^{-1} \leq T(\psi)$. We then have 
\begin{align*}
\int_{\R^3}|j^p|^2\rho^{-1} &\leq \inf \Big\{ T(\psi)  \Big| \,\psi \in W_N,\psi\mapsto(\rho,j^p) \Big\}\\ 
&\leq\inf \{ (\psi,H_0 \psi)_{L^2}| \psi \in W_N,\psi\mapsto(\rho,j^p) \} =Q(\rho,j^p).\,\,\,\blacksquare
\end{align*}
\\
\indent The situation is now as follows. The set of $v$-representable density pairs, $A_N$, is a proper subset of the $N$-representable 
density pairs, $Y_N$. The Hohenberg-Kohn functional $F_{HK}$, defined on $A_N$, has been extended to the Levy-Lieb-type functional 
$Q(\rho,j^p)$, which is defined on $Y_N$. Combining the variational principle of CDFT with Theorem \ref{thm3.4}, one has 
for $(v,A)\in V_N$,
\begin{align*}
e_0(v,A) = \min \Big\{ Q(\rho,j^p) + 2\int_{\R^3} j^p\cdot A + \int_{\R^3} \rho(v + |A|^2)\Big|(\rho,j^p)\in A_N \Big\}.
\end{align*}
The admissible set $A_N$ over which the minimization is performed can be exchanged by $Y_N$ if the minimum is 
replaced by infimum. Note that $Q(\rho,j^p) + 2\int_{\R^3} j^p\cdot A + \int_{\R^3} \rho(v + |A|^2)$ remains finite since we require 
$v\in L^{3/2}(\R^3) + L^\infty(\R^3)$ and $A^k\in L^\infty(\R^3)$.
\begin{theorem}
For $v\in L^{3/2}(\R^3) + L^\infty(\R^3)$ and $A^k\in L^\infty(\R^3)$, $$e_0(v,A) = \inf \Big\{ Q(\rho,j^p) + 
2\int_{\R^3} j^p\cdot A + \int_{\R^3} \rho(v + |A|^2)\Big|(\rho,j^p)\in Y_N \Big\}.$$
\label{thm11}
\end{theorem}
{\it Proof.} Fix $(\rho,j^p)\in Y_N$. Then 
\begin{align*}
e_0(v,A) &= \inf \Big\{ (\psi,H(v,A)\psi)_{L^2} \Big| \psi\in W_N \Big\} \\
& \leq \inf \Big\{ (\psi,H(v,A)\psi)_{L^2} \Big| \psi\in W_N,\rho_\psi=\rho,j_\psi^p=j^p \Big\}\\
& = Q(\rho,j^p) + 2\int_{\R^3} j^p\cdot A + \int_{\R^3} \rho(v + |A|^2).
\end{align*}
Since $(\rho,j^p)\in Y_N$ was arbitrary, we have 
\begin{align*}
e_0(v,A) \leq \inf \Big\{Q(\rho,j^p) + 2\int_{\R^3} j^p\cdot A + \int_{\R^3} \rho(v + |A|^2)\Big|(\rho,j^p)\in Y_N \Big\}.
\end{align*}
For the reverse inequality, let $\{\psi_k \}_{k=1}^\infty\subset W_N$ be a minimizing sequence for $e_0(v,A)$, i.e., 
$e_0(v,A) + \frac{1}{k}> (\psi_k,H(v,A)\psi_k)_{L^2}$. Put $\rho_k = \rho_{\psi_k}$ and $j_k^p= j_{\psi_k}^p$, then 
\begin{align*}
e_0(v,A) + \frac{1}{k} &> (\psi_k,H_0\psi_k)_{L^2} + 2\int_{\R^3}j_k^p\cdot A  + \int_{\R^3} \rho_k(v+|A|^2)\\
&\geq Q(\rho_k,j_k^p) + 2\int_{\R^3}j_k^p\cdot A  + \int_{\R^3} \rho_k(v+|A|^2)\\
&\geq \inf \Big\{Q(\rho,j^p) + 2\int_{\R^3} j^p\cdot A + \int_{\R^3} \rho(v + |A|^2)\Big|(\rho,j^p)\in  Y_N \Big\}.\,\,\,\blacksquare
\end{align*}
\\
\indent The admissible set over which the minimization is performed can be extended even 
further. First a definition.\\
\\
\textbf{Definition.} $X= \{(\rho,j^p)\vert \rho\in L^1(\R^3)\cap L^3(\R^3), j^p\in (L^1(\R^3))^3 \}$. \\
\\
\indent Next, for $(\rho,j^p)\in X$, define a functional $\tilde{Q}(\rho,j^p)$ given by 
\begin{align*}
\tilde{Q}(\rho,j^p) &= Q(\rho,j^p)\,\,\, \text{if $(\rho,j^p)\in Y_N$},\\
&= \infty, \,\,\, \text{otherwise}.
\end{align*}
The energy $e_0(v,A)$ can be computed using $\tilde{Q}$ on 
$X$. This is implied by the following argument. Let 
\begin{align*}
\tilde{e}(v,A) = \inf\left\{\tilde{Q}(\rho,j^p)+ 2\int_{\R^3}j^p\cdot A +
\int_{\R^3} \rho(v + |A|^2)\Big| \,(\rho,j^p) \in X \right\}.
\end{align*}
One directly has $\tilde{e}(v,A)\leq e_0(v,A)$, since $\tilde{Q}(\rho,j^p) = Q(\rho,j^p)$ if $(\rho,j^p)\in Y_N$. On the other hand, since 
$\tilde{Q}(\rho,j^p)=\infty$ if $(\rho,j^p)\notin Y_N$, we have $\tilde{e}(v,A) = e_0(v,A)$. Thus
\begin{theorem}
For $v\in L^{3/2}(\R^3) + L^\infty(\R^3)$ and $A^k\in L^\infty(\R^3)$,
\begin{align*}
e_0(v,A) 
= \inf\left\{\tilde{Q}(\rho,j^p)+ 2\int_{\R^3}j^p\cdot A +
\int_{\R^3} \rho(v + |A|^2)\Big|\, (\rho,j^p) \in X \right\}.
\end{align*}
\label{mathcalE}
\end{theorem}

\subsection{C. Convex envelope of $Q(\rho,j^p)$}
So far the variational principle of CDFT has been replaced by the following optimization problem
\begin{align*}
e_0(v,A) = \inf\left\{ Q(\rho,j^p) + 2\int_{\R^3}j^p\cdot A + \int_{\R^3}\rho\left( v +|A|^2\right)\Big\vert (\rho,j^p) \in Y_N\right\}.
\end{align*}
Note that $Q(\rho,j^p)$ could be exchanged by $\tilde{Q}(\rho,j^p)$ and the admissible set $Y_N$ extended to $X$. 
However, just as Lieb has demonstrated that $F_{LL}(\rho)$ is not convex (Theorem 3.4 in \cite{Lieb83}), 
the same is also true about $Q(\rho,j^p)$. A proof of this fact as well as a proof of (ii) in Proposition 4 now follows.
\begin{proposition}
$Q(\rho,j^p)$ is not a convex functional.
\label{PropNotCon}
\end{proposition}
{\it Proof of Proposition 8 and Proposition 4 (ii).} Choose $v(x)$ as in the proof of Theorem 3.4 in \cite{Lieb83} such that it has $M=2L+1$ ground-states $\psi_k$. Set 
$\rho_k = \rho_{\psi_k}$ and 
$j_k^p = j_{\psi_k}^p$ for $k=1,2,\dots, M$ and note that for all $k$,
\begin{align}
e_0(v,0) = Q(\rho_k,j_k^p) + \int_{\R^3} \rho_k v.
\label{Nacka1}
\end{align}
Let $\tilde{\rho}= \frac{1}{M}\sum_{k=1}^M \rho_k$ and $\tilde{j}^p= \frac{1}{M}\sum_{k=1}^M j_k^p$. By definition, 
$F_{LL}(\tilde{\rho})\leq Q(\tilde{\rho},\tilde{j}^p)$. One has
\begin{align*}
e_0(v,0)< F_{LL}(\tilde{\rho})+ \int_{\R^3}\tilde{\rho}v \leq Q(\tilde{\rho},\tilde{j}^p) + \int_{\R^3} \tilde{\rho} v, 
\end{align*}
where the first strict inequality follows by Theorem 3.4 of \cite{Lieb83} ($\tilde{\rho}$ cannot be a ground-state density of this $v(x)$). 
Using \eqref{Nacka1}, we obtain
\begin{align*}
\frac{1}{M}\sum_{k=1}^M Q(\rho_k,j_k^p) < Q(\tilde{\rho},\tilde{j}^p),
\end{align*}
which shows that $Q(\rho,j^p)$ is not convex.

For the proof of part (ii) in Proposition 4, assume that $\tilde{\rho}$ and $\tilde{j}^p$ are the ground-state densities of some other potential pair 
$(\tilde{v},\tilde{A})$, then
\begin{align*}
e_0(\tilde{v},\tilde{A}) &= Q(\tilde{\rho},\tilde{j}^p) +2 \int_{\R^3}\tilde{j}^p\cdot \tilde{A} + \int_{\R^3} \tilde{\rho}
(\tilde{v} + |\tilde{A}|^2) \\
&>  \frac{1}{M}\sum_{k=1}^M \left( Q(\rho_k,j_k^p)  +2 \int_{\R^3}j_k^p\cdot \tilde{A} + \int_{\R^3} \rho_k (\tilde{v} + |\tilde{A}|^2)\right).
\end{align*}
This gives that for at least one $k$,
\begin{align*}
e_0(\tilde{v},\tilde{A}) > Q(\rho_k,j_k^p)  +2 \int_{\R^3}j_k^p\cdot \tilde{A} + \int_{\R^3} \rho_k (\tilde{v} + |\tilde{A}|^2).
\end{align*}
But this is a contradiction and hence $A_N \subsetneq                Y_N$. $\,\,\,\blacksquare$\\

The next step will be to obtain a convex and universal density functional, denoted $F(\rho,j^p)$. For that purpose 
the Legendre transform will be used. The functional $F(\rho,j^p)$ will be defined 
on the whole space $X$. (Recall that 
$X$ is the space of those $(\rho,j^p)$ such that $\rho\in L^1(\R^3)\cap L^3(\R^3)$ and $j^p\in (L^1(\R^3))^3$.)\\
\\
\textbf{Definition.} The convex functional $F(\rho,j^p)$, defined on $X$, is given by
\begin{align*}
F(\rho,j^p) = \sup \Big\{ e_0(v,A) - 2\int_{\R^3} j^p\cdot A - \int_{\R^3} \rho(v+ |A|^2)\Big| v\in L^{3/2} + L^\infty, A^k\in L^\infty \Big\}.
\end{align*}
\\
\textbf{Remarks.} (i) Since $F$ is the supremum over $v$ and $A$ of linear functionals in $\rho$ and $j^p$, it is convex.

(ii) Furthermore, from the fact that 
$e_0(v,A)- 2\int_{\R^3} j^p\cdot A - \int_{\R^3} \rho(v+ |A|^2) \leq Q(\rho,j^p)$ for $(\rho,j^p)\in Y_N$, it follows that 
$F(\rho,j^p)\leq Q(\rho,j^p)$ for all $(\rho,j^p)\in Y_N$.\\

The functional $F(\rho,j^p)$ can be used to compute the ground-state energy, which follows from a direct generalization of Lieb's proof for the 
functional 
\[\sup \Big\{ e_0(v,0) - \int_{\R^3} \rho v\Big| v\in L^{3/2} + L^\infty \Big\}.\] 
One may minimize $F(\rho,j^p) +2 
\int_{\R^3} j^p\cdot A    +  \int_{\R^3}\rho (v +  |A|^2 )$ on either $Y_N$ or $X$.
\begin{theorem} 
\begin{align*}
e_0(v,A) &= \inf\Big\{F(\rho,j^p) +2 \int_{\R^3} j^p\cdot A    +  \int_{\R^3}\rho (v +  |A|^2 ) \Big\vert (\rho,j^p)
\in X\Big\}\\
 &= \inf\Big\{F(\rho,j^p) +2 \int_{\R^3} j^p\cdot A    +  \int_{\R^3}\rho (v +  |A|^2 ) \Big\vert (\rho,j^p)
\in Y_N\Big\}.
\end{align*}
\end{theorem}
{\it Proof.} Denote the first expression of $e_0(v,A)$ as $M^-(v,A)$ and the second one as $M^+(v,A)$. Note that $M^-(v,A)\leq M^+(v,A)$. 
Now, fix $v_0\in L^{3/2} + L^\infty$ and $A_0^k\in L^\infty$. By the definition of $F(\rho,j^p)$, we have for $(\rho,j^p)\in X$,
\begin{align*}
F(\rho,j^p) \geq e_0(v_0,A_0) -2 \int_{\R^3} j^p\cdot A_0    -  \int_{\R^3}\rho (v_0 +  |A_0|^2 ) = F_0(\rho,j^p),
\end{align*}
where the last equality is a definition. Thus
\begin{align*}
M^-(v_0,A_0) &\geq \inf\Big\{F_0(\rho,j^p) +2 \int_{\R^3} j^p\cdot A_0 + \int_{\R^3}\rho ( v_0 + |A_0|^2) 
\Big\vert (\rho,j^p)\in X  \Big\} \\&= e_0(v_0,A_0). 
\end{align*}
But since $v_0\in L^{3/2} + L^\infty$ and $A_0^k\in L^\infty$ was arbitrary, we obtain $M^-(v,A) \geq e_0(v,A)$.

On the other hand, for $(\rho,j^p)\in Y_N$ we have that $F(\rho,j^p)\leq Q(\rho,j^p)$, and consequently
\begin{align*}
M^+(v,A) \leq \inf\Big\{Q(\rho,j^p) +2 \int_{\R^3} j^p\cdot A   + \int_{\R^3} \rho\left(v +|A|^2 \right)\Big\vert (\rho,j^p)\in 
Y_N \Big\} = e_0(v,A).
\end{align*}
Thus $e_0(v,A) \leq M^-(v,A) \leq M^+(v,A) \leq e_0(v,A)$. $\,\,\,\blacksquare$\\

As the reader may recall, $Q(\rho,j^p) = F_{HK}(\rho,j^p)$ on $A_N$. In fact, $F(\rho,j^p)= Q(\rho,j^p) = F_{HK}(\rho,j^p)$ on $A_N$, since
\begin{proposition}
If $(\rho,j^p)\in A_N$, $F(\rho,j^p) = Q(\rho,j^p)$.
\end{proposition}
{\it Proof.} Assume $(\rho,j^p)\in A_N$, then $F_{HK}(\rho,j^p) = Q(\rho,j^p)$. For some $v$ and $A$, 
\begin{align*}
e_0(v,A) = Q(\rho,j^p) + 2\int_{\R^3}j^p\cdot A + \int_{\R^3} \rho(v + |A|^2).
\end{align*}
Conversely, using Theorem 9, 
\begin{align*}
e_0(v,A) \leq F(\rho,j^p) + 2\int_{\R^3}j^p\cdot A + \int_{\R^3} \rho(v + |A|^2).
\end{align*}
Thus $F(\rho,j^p) \geq Q(\rho,j^p)$. However, since the reverse inequality also holds, $F(\rho,j^p) = Q(\rho,j^p)$. $\,\,\,\blacksquare$\\

The main result of this section is the following:
\begin{theorem}
$F$ is the convex envelope of $Q$.
\label{ConvexEn}
\end{theorem}

Before proving Theorem \ref{ConvexEn}, some preparation is required. First define\\
\\
\textbf{Definitions.} (i) A functional, $f$, is weakly lower semi continuous (weakly l.s.c.) if 
\[f(\phi) \leq \liminf_{k\rightarrow \infty} f(\phi_k)\] 
when $\{\phi_k\}$ converges weakly to $\phi$.

(ii) Let $Z$ be a normed space and let $f:D\rightarrow \mathbb{R}$, where $D\subset Z$, and set
$$ \Lambda_{f,D} = \{g\,|\text{ $g$ is weakly l.s.c. and convex, and $g(\phi)\leq f(\phi)$ for all $\phi\in D$} \}.  $$
The convex envelope on $Z$ of the functional $f$ is 
then defined to be
\begin{align*}
\text{CE}\,f(\phi) = \sup\{g(\phi)\,|\,g\in\Lambda_{f,D}  \}.
\end{align*}

(iii) If $Z$ is a normed space we let $Z^*$ denote the dual space of $Z$, which is the space of all bounded linear functionals on $Z$. Moreover, 
the dual pairing between an element $z\in Z$ and $z^*\in Z^*$ will be denoted $\langle z,z^*\rangle_{Z,Z^*}$. If $Z=L^p(\R^3)$, then $Z^*=L^q(\R^3)$ with $1/p+1/q =1$, and $\langle z,z^*\rangle_{Z,Z^*}
= \int_{\R^3} z(x) z^*(x)dx$. In particular, 
\[X^* =\{(v',A')|v'\in L^{3/2}(\R^3) + L^\infty(\R^3),A'\in (L^\infty(\R^3))^3  \}.\]
\begin{proposition}
$F$ is weakly lower semi continuous.
\end{proposition}
{\it Proof.} The proof of this fact is standard, but is included for the sake of completeness. 
Since $F$ is convex, it suffices to show that $F$ is lower semi continuous in norm. 
For any $\lambda\in \mathbb{R}$, define the set
\begin{align*}
K_\lambda &= \{(\rho,j^p) | F(\rho,j^p) \leq \lambda  \} \\
&= \Big\{(\rho,j^p) \Big\vert e_0(v,A) -2 \int_{\R^3} j^p\cdot A - \int_{\R^3} \rho\left(v + |A|^2\right) \leq \lambda, 
 \forall\,\, (v,A)\in X^* \Big\}.
\end{align*}  
Now, assume that $\{\rho_n,j_n^p \}\subset K_\lambda$ and that $\rho_n\rightarrow\rho$ in $L^1$- and $L^3$-norm and that each component of 
$j_n^p$ converges to the respective component of $j^p$ in $L^1$-norm. Then for each $v\in L^{3/2} + L^\infty$ and $A^k\in L^\infty$,
\begin{align*}
\lambda &\geq \lim_n\left(e_0(v,A)-2 \int_{\R^3} j_n^p\cdot A - \int_{\R^3} \rho_n\left(v + |A|^2\right)  \right) \\
&=e_0(v,A) -2 \int_{\R^3} j^p\cdot A - \int_{\R^3} \rho\left(v + |A|^2\right),
\end{align*} 
which follows from the fact that norm convergence implies weak convergence. This shows that $K_\lambda$ is norm closed and that $F$ is 
norm l.s.c. $\,\,\,\blacksquare$\\

Now, let $f$ be a convex functional defined on a convex subset $D\subset Z$ of a normed space 
$Z$. Then the Legendre transform of $f$, denoted $f^*$, defined on the set
\begin{align*}
D^* = \{z^* \in Z^*| \sup_{z\in D}\{\langle z,z^*\rangle_{Z,Z^*}-f(z) \}<\infty \},
\end{align*}
is given by
\begin{align*}
f^*(z^*) = \sup\{\langle z,z^*\rangle_{Z,Z^*}-f(z)|z\in D \}.
\end{align*}
{\it Proof of Theorem \ref{ConvexEn}.} Let $f= \text{CE}\, Q$ and $D=  Y_N$. Note that $-\text{CE}\,Q\leq 0$, since $0\in \Lambda_{Q,Y_N}$. Then, 
for $(v',A')\in X^*$,
\begin{align*}
f^*(v',A') = \sup\left\{\int_{\R^3}\rho v' + \int_{\R^3}j^p \cdot A' - \text{CE}\, Q(\rho,j^p)\Big| (\rho,j^p)\in  Y_N \right\}.
\end{align*}
By the definition of the convex envelope, it follows that $\text{CE}\,Q(\rho,j^p)\leq Q(\rho,j^p)$ on $Y_N$. This gives
\begin{align*}
f^*(v',A') &= \sup\left\{ -\text{CE}\,Q(\rho,j^p) + \int_{\R^3}\rho v' + \int_{\R^3} j^p\cdot A'\Big| (\rho,j^p)\in Y_N \right\}\\
&\geq -\inf\left\{Q(\rho,j^p) - \int_{\R^3}\rho v' - \int_{\R^3} j^p\cdot A'\Big| (\rho,j^p)\in  Y_N \right\} \\&= -e_0(-v'-|A'/2|^2,-A'/2).
\end{align*}
By taking the Legendre transform one more time, one obtains for $(\rho.j^p)\in X$
\begin{align*}
(f^*)^*(\rho,j^p) &= \sup\left\{ \int_{\R^3}\rho v' + \int_{\R^3} j^p\cdot A'  - f^*(v',A')\Big| (v',A')\in X^* \right\}\\
&\leq \sup\left\{e_0(-v'-|A'/2|^2,-A'/2) + \int_{\R^3} \rho v' + \int_{\R^3} j^p\cdot A'\Big\vert (v',A')\in X^*     \right\} \\
&= \sup\left\{ e_0(v,A) -2 \int_{\R^3} j^p\cdot A - \int_{\R^3}\rho\left(v + |A|^2 \right)  \Big\vert (v,A)\in  X^*  \right\} = F(\rho,j^p),
\end{align*}
where $v= -v' - |A'/2|^2\in L^{3/2}(\R^3) + L^\infty(\R^3)$ and $A =- A'/2\in (L^\infty(\R^3))^3$. We may then conclude that, 
for $(\rho,j^p)\in X$,
\begin{align*}
(f^*)^*(\rho,j^p) \leq F(\rho,j^p).
\end{align*}

Now, from an infinite dimensional extension of Fenchel's theorem it follows that if the original functional is convex and 
weakly lower semi continuous, then the double Legendre transform of the functional equals the functional itself \cite{Lieb83}. Thus for $f = \text{CE}\, Q$ we obtain
\begin{align*}
\text{CE}\, Q(\rho,j^p) = f(\rho,j^p) = (f^*)^*(\rho,j^p)\leq  F(\rho,j^p).
\end{align*}
Conversely, since $F$ is convex, weakly lower semi continuous and is bounded above by $Q$, i.e. $F\in \Lambda_{Q,Y_N}$, we have that 
\begin{align*}
F(\rho,j^p) \leq \sup \{ f(\rho,j^p)\vert f\in \Lambda_{Q,Y_N} \} = \text{CE}\, Q(\rho,j^p).
\end{align*}
It then follows that for all $(\rho,j^p)\in X$, $\text{CE}\, Q(\rho,j^p) = F(\rho,j^p)$. $\,\,\,\blacksquare$\\

Since $F(\rho,j^p)$ is convex, one may seek to obtain a connection between a set of Euler-Lagrange equations and the minimization of 
$F(\rho,j) + 2\int_{\R^3} j^p\cdot A + \int_{\R^3} \rho(v+|A|^2)$ on $Y_N$. Let $Z$ be a normed space and $f$ a real-valued 
functional on $Z$, $f:Z\rightarrow \R$. If $f$ is convex on $Z$, given $z_0\in Z$, there exists $z^*\in Z^*$, not 
necessarily unique, such that
\begin{align*}
f(z) \geq f(z_0) + \langle z - z_0, z^*\rangle_{Z,Z^*}
\end{align*}
holds for all $z\in Z$. We now introduce the concept of Fr\'echet differentiability and Fr\'echet derivative. Let $f:Z\rightarrow \mathbb{R}$ 
be defined on an open domain $D_f\subset Z$. If, for a fixed $z\in Z$ and for each $h \in Z$, there exists $\delta f(z;h)\in \mathbb{R}$ that 
is linear and continuous with respect to $h$ such that
\begin{align*}
\lim_{||h||_Z\rightarrow 0} \frac{|f(z + h) - f(z) - \delta f(z;h)  |}{||h||_Z} = 0,
\end{align*} 
then $f$ is said to be Fr\'echet differentiable at $z$ and $\delta f(z;h)$ is said to be the Fr\'echet differential 
of $f$ at $z$ with increment $h$. The Fr\'echet differential is unique, and if it exists then
\begin{align*}
\lim_{\alpha\rightarrow 0} \frac{f(z + \alpha h) - f(z)}{\alpha}
\end{align*} 
exists and equals the Fr\'echet differential. We write $\delta f(z;h) =\langle h,f'(z)\rangle_{Z,Z^*}$ and call $f'$ the Fr\'echet 
derivative of $f$. Note that if $f'(z_0)$ exists, we have for all $z$
\begin{align*}
f(z) \geq f(z_0) + \langle z - z_0, z^*\rangle_{Z,Z^*},
\end{align*}
where $z^*=f'(z_0)$ is unique. For a functional $f(z_1,z_2)$, $f_{z_k}'$ will be used to denote partial derivative. 
We are now ready to formulate and prove
\begin{theorem}
Assume that $F_\rho'(\rho_0,j_0^p)$ and $F_{j^p}'(\rho_0,j_0^p)$ exist and $\int_{\R^3}\rho_0=N$ and that 
\begin{align*}
F_\rho'(\rho_0,j_0^p) + v + |A|^2 +\mu_0 &=0,\\
F_{j^p}'(\rho_0,j_0^p) + 2A  &=0,
\end{align*}
a.e. for some $\mu_0\in \R$, $v\in L^{3/2}(\R^3) + L^\infty(\R^3)$ and $A \in(L^\infty(\R^3))^3$. Then $(\rho_0,j_0^p)$ minimizes 
\begin{align*}
\inf\Big\{ F(\rho,j^p)+2\int_{\R^3} j^p\cdot A + \int_{\R^3}\rho( v + |A|^2)\Big|(\rho,j^p)\in Y_N \Big\}.
\end{align*}
If in addition, $F(\rho_0,j_0^p) = Q(\rho_0,j_0^p)$, then $(\rho_0,j_0^p)\in A_N$.
\label{thmB}
\end{theorem}
{\it Proof.} Set $w_1=-F_\rho'(\rho_0,j_0^p)\in L^{3/2}(\R^3) + L^\infty(\R^3)$ and $w_2=-F_{j^p}'(\rho_0,j_0^p)\in(L^\infty(\R^3))^3$. 
Since $F'$ exists at $(\rho_0,j_0^p)$ and $F$ is convex, we have for $(\rho,j^p)\in Y_N$,
\begin{align*}
F(\rho,j^p) \geq F(\rho_0,j_0^p) + \int_{\R^3} (j_0^p-j^p)\cdot w_2 + \int_{\R^3}(\rho_0-\rho) w_1.
\end{align*}
By assumption, $w_1=v + |A|^2  + \mu_0$ and $w_2 =2A$ a.e. Since $\int_{\R^3}\rho_0=N$, 
\begin{align*}
F(\rho,j^p) \geq F(\rho_0,j_0^p) +2\int_{\R^3} (j_0^p-j^p)\cdot A + \int_{\R^3}(\rho_0-\rho) (v+|A|^2) + \mu_0( N -\int_{\R^3}\rho).
\end{align*}
However, for any $(\rho,j^p)\in Y_N$, $\int_{\R^3}\rho =N$, and hence the conclusion follows.

For the second part, assume $F(\rho_0,j_0^p) = Q(\rho_0,j_0^p)$. Using $Q(\rho,j^p)\geq F(\rho,j^p)$, we obtain
\begin{align*}
Q(\rho,j^p)\geq F(\rho,j^p)&\geq F(\rho_0,j_0^p) - \int_{\R^3} (\rho-\rho_0)w_1 - \int_{\R^3} (j^p-j_0^p)\cdot w_2 \\
&= Q(\rho_0,j_0^p)- \int_{\R^3} (\rho-\rho_0)w_1 - \int_{\R^3} (j^p-j_0^p)\cdot w_2.
\end{align*}
If we define $\tilde{Q}(\rho,j^p)= Q(\rho_0,j_0^p)- \int_{\R^3} (\rho-\rho_0)w_1 - \int_{\R^3} (j^p-j_0^p)\cdot w_2$, we have $\tilde{Q}(\rho,j^p)\leq 
Q(\rho,j^p)$. Now,
\begin{align*}
&Q(\rho_0,j_0^p) +\int_{\R^3} j_0^p\cdot w_2 + \int_{\R^3} \rho_0 w_1 = \inf\Big\{ \tilde{Q}(\rho,j^p)+\int_{\R^3} j^p\cdot w_2 + \int_{\R^3} \rho w_1\Big| (\rho,j^p)\in Y_N\Big\}\\
&\leq e_0\Big(w_1 -\frac{|w_2|}{4}^2,\frac{w_2}{2}\Big)\leq Q(\rho_0,j_0^p) +\int_{\R^3} j_0^p\cdot w_2 + \int_{\R^3} \rho_0 w_1.
\end{align*}
By setting $w_1=v + |A|^2\in L^{3/2}(\R^3) + L^\infty(\R^3)$ and $w_2=2A\in (L^\infty(\R^3))^3$, it follows that
\begin{align*}
e_0(v,A) = Q(\rho_0,j_0^p) +2\int j_0^p\cdot A + \int \rho_0(v+|A|^2).
\end{align*}
From Theorem 5 (i), we know that there exists a $\psi_0\in W_N$ such that $Q(\rho_0,j_0^p) = (\psi_0,H_0\psi_0)_{L^2}$ and 
$\psi_0\mapsto (\rho_0,j_0^p)$. But then
\begin{align*}
e_0(v,A) = (\psi_0,H_0\psi_0)_{L^2} - 2\int_{\R^3} j_0^p\cdot + \int_{\R^3} \rho_0(v+|A|^2) = (\psi_0,H(v,A)\psi_0)_{L^2},
\end{align*}
which shows that $(\rho_0,j_0^p)\in A_N$.$\,\,\,\blacksquare$\\

The last order of business in this section will be to obtain a lower bound for $F$ on $X$. The motivation is the following. From Theorem 3.8 in 
\cite{Lieb83}, we have 
\begin{align*}
F_c(\rho) = \text{CE}\, F_{LL}(\rho) & \geq \int_{\R^3} (\nabla \rho^{1/2})^2,\,\,\,\text{if $\rho\in I_N$},\\
    &\geq \infty,\,\,\,\text{otherwise}.
\end{align*}
We shall now take convex combinations of the two convex functionals $\rho\mapsto\int_{\R^3} (\nabla \rho^{1/2})^2$ and 
$(\rho,j^p)\mapsto N^2\int_{\R^3} |j^p|^2\rho^{-1}$ and use Theorem \ref{ConvexEn} to obtain
\begin{theorem}
Define for $(\rho,j^p)\in X$ and $0\leq \lambda\leq 1$
\begin{align*}
J_\lambda(\rho,j^p) &= \lambda \int_{\R^3}(\nabla \rho(x)^{1/2})^2  +(1-\lambda)\int_{\R^3} |j^p(x)|^2\rho(x)^{-1}, \,\,\, \text{if $(\rho,j^p)\in Y_N$},\\
&= \infty, \,\,\, \text{otherwise}.
\end{align*}
Then $J_\lambda(\rho,j^p)\leq F(\rho,j^p)$ for $(\rho,j^p)\in X$ and $0\leq \lambda\leq 1$.
\label{GEN3.8}
\end{theorem}
{\it Proof.} From \cite{Lieb83} we have that $\rho\mapsto \int_{\R^3} (\nabla\rho^{1/2})^2$ is convex and bounded above by $T(\psi)$ for 
$\rho_\psi =\rho$, $\psi\in W_N$. From Proposition \ref{PROP1} and Theorem \ref{thm3.4}, we can then conclude that 
$J_\lambda$ is convex and $J_\lambda\leq Q$ on $Y_N$. We now want to show that $J_\lambda$ is weakly l.s.c. (since $J_\lambda$ is convex we will show that 
it is norm-l.s.c) so we can conclude that 
$J_\lambda\leq \text{CE}\, Q=F$. 

Let $\rho_n\rightarrow \rho$ in $L^1(\R^3)\cap L^3(\R^3)$-norm and $j_n^p\rightarrow j^p$ in $(L^1(\R^3))^3$-norm. We want to show that 
$C_\lambda=\liminf_{n\rightarrow\infty}J_\lambda(\rho_n,j_n^p)\geq J_\lambda(\rho,j^p)$. If $C_\lambda=\infty$ we are done, so 
we will assume that $C_\lambda<\infty$. 

Note that the case $\lambda =1$ follows from Theorem 3.8 in \cite{Lieb83}. It then suffices to show the result for $\lambda=0$. 
By the same argument as in the proof of Theorem 3.8 in \cite{Lieb83}, assume $\rho\in I_N$. (If $\rho<0$ on a set 
of positive measure, then $\rho_n<0$ and $J_0(\rho_n,j_n^p)=\infty$ for sufficiently large $n$. Similarly we have that $\int_{\R^3}\rho\neq N$ 
gives $J_0(\rho_n,j_n^p)=\infty$ for sufficiently large $n$.) Since $C_0<\infty$, $(\rho_n,j_n^p)\in Y_N$. 
Set $g_n=j_n^p/\rho_n^{1/2}$. Then $\{g_n\}$ (or at least a subsequence of $\{g_n\}$) is bounded in $L^2(\R^3)^3$, and 
by the Banach-Alaoglu theorem there exists a $g\in L^2(\R^3)^3$ and 
a subsequence $\{g_{n_k}\}$ such that $g_{n_k}\rightharpoonup g$ in $L^2(\R^3)^3$. 

The next step is to show that $g=j^p/\rho^{1/2}$ a.e. First note since $\rho_{n_k}\rightarrow \rho\geq 0$ in 
$L^1(\R^3)$-norm, there exists a subsequence, which we continue to denote $\{ \rho_{n_k} \}$, and a non-negative $F\in L^1(\R^3)$ such that 
$\rho_{n_k}(x)\leq F(x)$ and $\rho_{n_k}(x)\rightarrow \rho(x)$ a.e.
From $$|\rho_{n_k}(x)^{1/2}-\rho(x)^{1/2}|^2\leq 2(\rho_{n_k}(x) + \rho(x))\leq 2(F(x) + \rho(x)),$$ we have by dominated convergence, $\rho_{n_k}^{1/2}\rightarrow \rho^{1/2}$ 
in $L^2(\R^3)$. Let $(u)_l$ denote the $l$:th component of the vector $u$. 
It then follows that $g_{n_k}\rho^{1/2} \rightarrow j^p$ in $L^1(\R^3)^3$, since for $l=1,2,3$,
\begin{align*}
\int_{\R^3} |(g_{n_k})_l \rho^{1/2} - (j^p)_l|   &\leq \int_{\R^3} |(g_{n_k})_l \rho^{1/2} - (g_{n_k})_l \rho_{n_k}^{1/2}|+
\int_{\R^3} |(g_{n_k})_l \rho_{n_k}^{1/2} - (j^p)_l|\\
&\leq   \left(\int_{\R^3} |(g_{n_k})_l|^2 \right)^{1/2} \left(\int_{\R^3}|\rho_{n_k}^{1/2}-\rho^{1/2}|^2   \right)^{1/2}
+\int_{\R^3} |(j_{n_k}^p)_l  - (j^p)_l|\rightarrow 0,
\end{align*}
as $k\rightarrow 0$, where we used that $\rho_{n_k}^{1/2}\rightarrow \rho^{1/2}$ in $L^2(\R^3)$ and $(j_{n_k}^p)_l\rightarrow (j^p)_l$ 
in $L^1(\R^3)$ for $l=1,2,3$. 

Now, let $M\subset \R^3$ be an arbitrary measurable set. Since $\rho^{1/2}\chi_M\in L^2(\R^3)$ and by the weak convergence of 
$g_{n_k}$ to $g$ in $L^2(\R^3)^3$, one obtains
\begin{align*}
\lim _{k\rightarrow \infty} \int_{\R^3} (g_{n_k})_l \rho^{1/2}\chi_M = \int_M (g)_l \rho^{1/2}.
\end{align*}
On the other hand, since $\chi_M\in L^\infty(\R^3)$ and norm-convergence 
implies weak-convergence, $\lim _{k\rightarrow \infty}
\int_{\R^3} (g_{n_k})_l \rho^{1/2}\chi_M  = \int_M (j^p)_l$. 
Then $\int_M (g)_l \rho^{1/2}=\int_M (j^p)_l$, which gives $g=j^p/\rho^{1/2}$ a.e.

Lastly, by the w.l.s.c. of the $L^2(\R^3)$-norm, 
\begin{align*}
C_0 =\liminf_{k\rightarrow\infty}\int_{\R^3} |j_{n_k}^p(x)|^2\rho_{n_k}(x)^{-1} &= \liminf_{k\rightarrow\infty}|| g_{n_k} ||_{L^2(\R^3)}^2\\
&\geq || g ||_{L^2(\R^3)}^2 = \int_{\R^3} |j^p(x)|^2\rho(x)^{-1} =J_0(\rho,j^p). \,\,\,\blacksquare
\end{align*}

\subsection{D. Upper and lower bounds for densities with vanishing vorticity}
The last issue to be addressed is when the density pair $(\rho,j^p)$ is restricted to the 
constraint $\nabla\times (j^p/\rho)=0$. The quantity $\nabla\times (j^p/\rho)$ is called the vorticity. 
We shall begin by constructing a determinantal 
wavefunction that yields a prescribed density pair $(\rho,j^p)$, i.e., finding 
a function in $W_N$ that is a determinant and that reproduces a given density pair 
$(\rho,j^p)\in  Y_N$. This can be achieved by a straightforward generalization of Theorem 1.2 of \cite{Lieb83} (see also \cite{LiebSchrader} 
where a determinantal construction is considered without the constraint $\nabla\times (j^p/\rho)=0$ for $N\neq 3$ but without 
an explicit upper bound for the kinetic energy). 
Define, as in ref. \cite{Lieb83}, a function on the real line given by
\begin{align*}
f(x_1) = \frac{2\pi}{N}\int_{-\infty}^{x_1}\int_{-\infty}^\infty\int_{-\infty}^\infty
\rho(s,x_2,x_3)\,ds\,dx_2\,dx_3.
\end{align*}
Note that $f(-\infty) =0$, $f(\infty) =2\pi$ and
\begin{align*}
\frac{df}{dx_1} = \frac{2\pi}{N} \int_{\R^2} \rho(x_1,x_2,x_3)\,dx_2\,dx_3. 
\end{align*}
To obtain a determinant $\psi_D$ that yields a given density pair $(\rho,j^p)\in  Y_N$, put 
\begin{align*}
\psi_D(x_1,\dots,x_N) = (N!)^{-1/2}\det [\phi_k(x_l)]_{k,l},
\end{align*}
where, for $k=0,1,\dots,N-1$,
\begin{align}
\phi_k(x) &= \left(\frac{\rho(x)}{N}\right)^{1/2} e^{i(kf(x_1)-M(x_1) + S(x))}.
\label{detCon1}
\end{align}
Note that $(\phi_k,\phi_l)_{L^2} = \delta_{kl}$. It is immediate that $\rho_{\psi_D}=\sum_{k=0}^{N-1}|\phi_k(x)|^2 = \rho$. Moreover, from the calculation
\begin{align*}
\text{Im}(\overline{\phi_k}\nabla\phi_k) = \frac{\rho}{N}
\left(\left(k\frac{df}{dx_1} -\frac{dM}{dx_1}\right)\hat{e}_x + \nabla S \right),
\end{align*}
it follows that
\begin{align}
j_{\psi_D}^p = \sum_{k=0}^{N-1} \text{Im}(\overline{\phi_k}\nabla\phi_k) &= \rho \nabla S +  
\left(\frac{\rho}{N}\frac{df}{dx_1} \sum_{k=0}^{N-1} k -\rho\frac{dM}{dx_1} \right)
\hat{e}_x \nonumber \\
&=\rho \nabla S + \rho\left(\frac{1}{2}(N-1)\frac{df}{dx_1} -\frac{dM}{dx_1} \right)\hat{e}_x.
\label{detCon2}
\end{align}
\begin{proposition}
Given $(\rho,j^p)\in  Y_N$ that fulfils $\nabla \times (j^p/\rho)=0$, there 
exists a determinant $\psi_D\in L^2(\R^{3N})$ such that $|| \psi_D||_{L^2} =1$ and 
\begin{align}
T(\psi_D)\leq  \left(1+(4\pi)^2 \frac{(N^2-1)}{12}  \right)\int_{\R^3} (\nabla\rho^{1/2})^2     +\int_{\R^3}|j^p|^2\rho^{-1}<\infty.
\label{DetKinBound}
\end{align}
\label{DetCon}
\end{proposition}
{\it Proof.} Take $\psi_D = (N!)^{-1/2}\det [\phi_k(x_l)]_{k,l}$, with $\phi_k$ as in \eqref{detCon1} for $k=0,1,\dots,N-1$. 
From \eqref{detCon2}, $j_{\psi_D}^p = j^p$ if $S$ and $M$ are chosen such that $\nabla S= j^p/\rho$ and 
$$M(x_1) = \frac{f(x_1)}{N}\sum_{k=0}^{N-1}k= \frac{1}{2}(N-1)f(x_1).$$ We are done if we can show \eqref{DetKinBound}. To that end, 
note that
\begin{align*}
|\nabla\phi_k|^2 = \frac{1}{N}\left( (\nabla\rho^{1/2})^2  +\rho 
\left(  \left(k\frac{df}{dx_1} -\frac{dM}{dx_1}\right)\hat{e}_x + \nabla S  \right)^2\right).
\end{align*}
The kinetic energy of $\psi_D$ satisfies
\begin{align}
T(\psi_D) &= \sum_{k=0}^{N-1}\int_{\R^3}|\nabla\phi_k|^2 \nonumber\\&= \int_{\R^3}(\nabla\rho^{1/2})^2 
+\left(\frac{1}{N}\sum_{k=0}^{N-1} k^2 - \frac{(N-1)}{4}^2\right)
\int_{\R^3} \rho\left(\frac{df}{dx_1}\right)^2 +\int_{\R^3}\rho|\nabla S|^2\nonumber\\
&= \int_{\R^3}(\nabla\rho^{1/2})^2 
+  \frac{(N^2-1)}{12} \int_{\R^3} \rho\left(\frac{df}{dx_1}\right)^2+\int_{\R^3}|j^p|^2\rho^{-1}.
\label{detCon3}
\end{align}
For the second term in the r.h.s. of \eqref{detCon3}, note that
\begin{align*}
\int_{\R^3} \rho(x)\left(\frac{df}{dx_1}\right)^2dx = \left(\frac{2\pi}{N}\right)^2\int_{\R}g(x_1)^6dx_1 ,
\end{align*}
where
\begin{align*}
g(x_1)^2 =  \int_{\R^2} \rho(x_1,x_2,x_3)\,dx_1\,dx_2.
\end{align*}
From \cite{Lieb83}, $g\in H^1(\R)$ and moreover 
\begin{align*}
g(x_1)^4 \leq 4\int_\R g(x_1)^2 dx_1 \int_\R \left|\frac{dg}{dx_1}\right|^2 dx_1 
\leq 4N\int_{\R^3}\left( \nabla \rho^{1/2}\right)^2dx.
\end{align*}
Thus, \eqref{detCon3} now gives
\begin{align*}
T(\psi_D) \leq \left(1+(4\pi)^2 \frac{(N^2-1)}{12}\right)\int_{\R^3} (\nabla\rho^{1/2})^2     +\int_{\R^3}|j^p|^2\rho^{-1}<\infty,
\end{align*}
where all terms are finite since $(\rho,j^p)\in Y_N$. $\,\,\,\blacksquare$\\
\\
\textbf{Remark.} Note that for $\psi_D$ chosen as in Proposition \ref{DetCon}, the exchange-correlation energy, $E_{xc}(\psi_D)$, 
does not depend on the paramagnetic current density. This can be seen from 
\begin{align*}
E_{xc}(\psi_D) &= -\frac{1}{2N^2} \int_{\R^3} \int_{\R^3} \frac{\rho(x)\rho(y)}{|x-y|} 
\left\vert \sum_{k=0}^{N-1} e^{ik(f(x_1)-f(y_1))}   \right\vert^2 dxdy\\
& = -\frac{1}{2N} \int_{\R^3} \int_{\R^3} \frac{\rho(x)\rho(y)}{|x-y|} F_N(f(x_1)-f(y_1) )dxdy,
\end{align*}
where $F_N(t)$ is the Fej\'er kernel, given by $F_N(t)= \sin^2\left(Nt/2 \right)/(N\sin^2\left(t /2\right))$.
\\
\begin{proposition}
For $(\rho,j^p)\in Y_N$ fulfilling $\nabla\times (j^p/\rho) =0$, we have
\begin{align*}
Q(\rho,j^p)\leq \left(1+(4\pi)^2 \frac{(N^2-1)}{12}  \right)\int_{\R^3} (\nabla\rho^{1/2})^2 +\int_{\R^3}|j^p|^2\rho^{-1}
+ \frac{1}{2} \int_{\R^3} \int_{\R^3} \frac{\rho(x)\rho(y)}{|x-y|}dxdy.
\end{align*}
\label{propD}
\end{proposition}
{\it Proof.} First note that 
$$(\psi,H_0\psi) = T(\psi) + E_{xc}(\psi) + 
\frac{1}{2} \int_{\R^3} \int_{\R^3} \frac{\rho_\psi(x)\rho_\psi(y)}{|x-y|}dxdy. $$
Now, given $(\rho,j^p)\in Y_N$ fulfilling $\nabla\times (j^p/\rho) =0$, 
there exits, by Proposition \ref{DetCon}, a determinantal wavefunction $\psi_D$ such that 
$\rho_{\psi_D}= \rho$ and $j_{\psi_D}^p = j^p$. We then have
\begin{align*}
Q(\rho,j^p)&\leq  \left(1+(4\pi)^2 \frac{(N^2-1)}{12}  \right)    \int_{\R^3} (\nabla\rho^{1/2})^2 +\int_{\R^3}|j^p|^2\rho^{-1}
+(\psi_D,\sum_{1\leq k<l\leq N}|x_k-x_l|^{-1}\psi_D)\\
&\leq \left(1+(4\pi)^2 \frac{(N^2-1)}{12}  \right)    \int_{\R^3} (\nabla\rho^{1/2})^2 +\int_{\R^3}|j^p|^2\rho^{-1}
+ \frac{1}{2} \int_{\R^3} \int_{\R^3} \frac{\rho(x)\rho(y)}{|x-y|}dxdy,
\end{align*}
where the last inequality follows from the fact that $E_{xc}(\psi_D)\leq 0$, since 
$\psi_D$ is a determinant. $\,\,\,\blacksquare$\\

We conclude this section by applying Proposition \ref{propD} and Theorem \ref{GEN3.8}. The following corollary gives 
both an upper and lower bound for $Q$ and $F$ in terms of $J_0(\rho,j^p) = \int_{\R^3}|j^p|\rho^{-1}$ and 
$J_1(\rho,j^p) =\int_{\R^3}(\nabla\rho^{1/2})^2$.
\begin{corollary}
Let $(\rho,j^p)\in Y_N$ be such that $\nabla\times(j^p/\rho) =0$. Then for $0\leq \lambda\leq 1$,
\begin{align*}
\lambda J_0(\rho,j^p) + (1-\lambda)J_1(\rho,j^p)&= J_\lambda(\rho,j^p) \leq F(\rho,j^p) \leq Q(\rho,j^p) \\ 
&\leq aN + (b+cN^2)J_1(\rho,j^p) + J_0(\rho,j^p),
 \end{align*}
 where $a= 4/( 3\sqrt{3}\pi)$, $b= 1-(4\pi)^2/12$ and $c=(4\pi)^2/12 + 4/( 3\sqrt{3}\pi)$.
 \label{corE}
\end{corollary}
{\it Proof.} The statement follows directly from Proposition \ref{propD} and Theorem \ref{GEN3.8} and the fact that
\begin{align*}
\frac{1}{2} \int_{\R^3} \int_{\R^3} \frac{\rho(x)\rho(y)}{|x-y|}dxdy &\leq C_1 ||\rho||_{L^{6/5}(\R^3)}^2 
\leq C_1 N^{3/2}||\rho||_{L^3(\R^3)}^{1/2}\\
&\leq C_1C_2 N^{3/2} J_1(\rho,j^p)^{1/2} \leq \frac{1}{\pi}\frac{4}{3\sqrt{3}} (N+ N^2 J_1(\rho,j^p)),
\end{align*}
where the Hardy-Littlewood-Sobolev inequality ($C_1 = 2(4/\pi^{1/2})^{2/3}/3$) and Sobolev's inequality for gradients 
($C_2 = 2/(3^{1/2}2^{1/3}\pi^{2/3})$) have been used \cite{LiebLoss}.$\,\,\,\blacksquare$\\

\section{IV. SUMMARY}
This paper has aimed at giving CDFT formulated with the paramagnetic current density a mathematically rigorous 
foundation. It has focused on defining and investigating density functionals that depend on the particle density and 
the paramagnetic current density. 
$N$-representable density pairs $(\rho,j^p)$ have been defined. A Hohenberg-Kohn 
functional, $F_{HK}(\rho,j^p)$, has been extended to a Levy-Lieb-type functional, denoted $Q(\rho,j^p)$, with the set of $N$-representable densities as domain. 
It has been proven that there exists a wavefunction $\psi_0$ such that 
$Q(\rho,j^p) = (\psi_0,H_0\psi_0)_{L^2}$ and $\rho_\psi=\rho$, $j_\psi^p =j^p$. Moreover, a universal and convex 
density functional $F(\rho,j^p)$ has been proven to exist such that it equals the convex envelope of $Q(\rho,j^p)$. On the set 
of $v$-representable densities, the functionals $F_{HK}$, $Q$ and $F$ all agree. Furthermore, a connection between the minimization of 
$F(\rho,j^p)$ and a set of Euler-Lagrange equations has been established.

For $N$-representable density pairs $(\rho,j^p)$ fulfilling $\nabla \times (j^p/\rho)=0$, both upper and lower bounds 
of $F$ and $Q$ in terms of convex functionals that are given explicitly have been obtained. 

\section{ACKNOWLEDGMENTS}
The author is very thankful to Michael Benedicks and Anders Szepessy for useful comments and discussions.

\bibliography{./references}

\end{document}